\documentclass[runningheads]{llncs}
\usepackage[
backend=biber,
style=numeric,
sorting=none
]{biblatex}

\usepackage[T1]{fontenc}
\usepackage{graphicx}
\begin{document}
\title{Reducing Student Distraction Through Fuzzy Logic Based Seating Arrangements}
%
%
\author{Garrett Olges\inst{1}\orcidID{0009-0009-4061-8239} \and
Kelly Cohen\inst{2}\orcidID{0000-0002-8655-1465}
}
\authorrunning{G. Olges}
%
\institute{University of Cincinnati, Cincinnati OH 45221, USA 
}
\maketitle              
\begin{abstract}
A crucial skill for primary school teachers is maintaining efficient classroom management. Teachers use classroom seating arrangements to help maintain this efficiency. However, developing classroom seating arrangements is both time-consuming and often non-optimal for distraction mitigation. Fuzzy logic-based approaches for the development of classroom seating arrangements can reduce development time and minimize classroom distraction. In this study, an original fuzzy logic-based software package named “CUB” is introduced and applied to a modern classroom using "cluster" seating arrangements. The combination of fuzzy inference systems, fuzzy c-means clustering, sequential, and iterative processes produce ready-to-use seating arrangements for the classroom in this study. The seating arrangements are compared with an existing set of seating arrangements to validate the results. The author’s findings show that CUB is successful in generating applicable seating arrangements with a small likelihood of replicating arrangements. The findings also suggest that fuzzy logic-based approaches may be successful in other styles of classroom arrangement.

\end{abstract}
\section{Introduction}
A significant component of the role of a primary school teacher is effective classroom management \cite{HOEKSTRA2023104016}. Developing classroom seating arrangements is one means by which teachers can increase the effectiveness of their classroom management. However, the process of developing classroom seating arrangements to improve classroom management efficiency is a complex task that involves a wide array of considerations. As a result, drafting new seating arrangements have the potential to require significant time from teachers as well as risk the production of non-optimal arrangements. The purpose of the current study is to minimize the time needed to develop seating arrangements and maximize classroom management efficiency through the incorporation of a soft computing methodology called "fuzzy logic." 
For the purposes of the current study, fuzzy logic is “a precise logic of imprecision, uncertainty and approximate reasoning” \cite{Zadeh2023}. The current study implements two fuzzy logic-based computing methods: fuzzy inference systems (FIS) and fuzzy c-means (FCM) clustering. Alongside these methods are algorithms used for classification and seat assignment. This combination of fuzzy methods and algorithms forms a complete software package that the author has named CUB. In addition to the previously mentioned goal, it is the goal of the current study to demonstrate the applicability of fuzzy logic principles through the documented use of CUB.

\section{Method}
\subsection{Overview}
CUB operates on a five-step process to generate seating arrangements. The process first begins by evaluating teacher-submitted surveys with a FIS. The FIS generates two coefficients for each student entry in the survey. These coefficients serve as inputs for the FCM clustering method. FCM clustering assigns each student entry one of three possible cluster labels with membership values detailing how attuned to each cluster the student entry is. A classification algorithm then uses the membership values to further classify the cluster labels, creating six possible category labels for each student entry. An assignment algorithm then generates a seating arrangement based on the category labels of each student entry whilst considering additional parameters set initially by the teacher. The sequencing of CUB is outlined in Figure 1. It is important to note that CUB assumes “groups/clusters” for the desired seating technique as the education facility involved in the study commonly implements this seating technique.

\begin{figure}
\includegraphics[width=\textwidth]{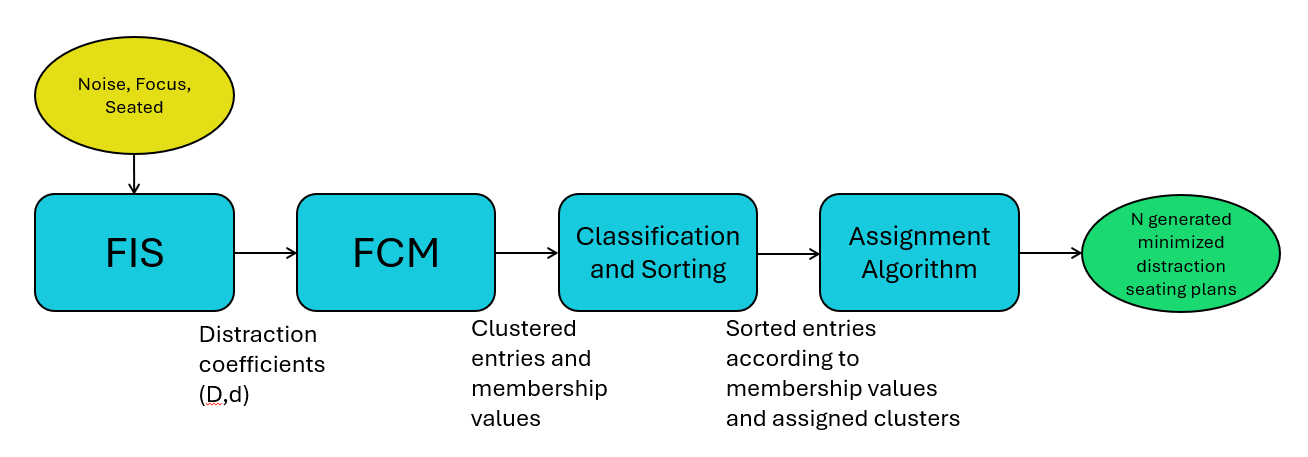}
\caption{Flow chart depicting the sequencing of CUB} \label{fig1}
\end{figure}

\subsection{Data Acquisition}
There are two types of information gathered from each educator: survey results and prior classroom seating arrangements. The survey results are the inputs for CUB while the prior classroom seating arrangements are used for validation.

\subsubsection{Surveys}
The teacher-submitted survey is a collection of English words and phrases answering three questions about each student in the classroom. Each question, along with the corresponding survey answers, is provided in Figure 2. A teacher would answer one set of three questions for each student. Upon conclusion of the survey, entries for each student are extracted to a machine-legible dataset.

\begin{figure}
\includegraphics[width=\textwidth]{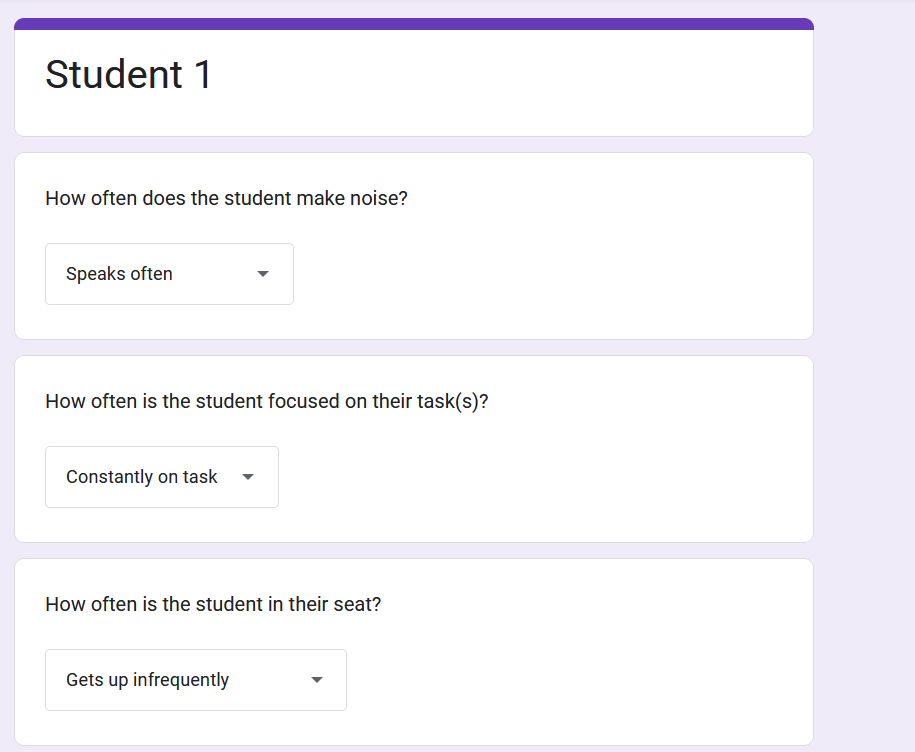}
\caption{Example of a survey for one student} \label{fig2}
\end{figure}

\subsubsection{Prior Classroom Seating Arrangements}
The current study was conducted over the course of one academic year. During the term, classroom seating arrangements were recorded and labeled according to their efficiency as each arrangement rotated out of use. These arrangements would be created by a teacher using their usual methods and employed for one month before being replaced by a new arrangement. These arrangements serve as a benchmark for evaluating CUB. Each prior arrangement was labeled “Poor, ““Acceptable,” or “Good.”

\subsection{Fuzzy Inference System}
A three-input two-output type-one Mamdani FIS is used to produce two coefficients from the survey data. The FIS consists of 441 rules, the total number of combinations between the three input questions. Each input of the FIS pertains to one of the answers in each student entry and are labeled accordingly (Noise, Focus, and Seated). The output of the FIS, \textit{D} and \textit{d}, refers to two key characteristics of each student entry. \textit{D} describes how prone a student is to being distracted, with higher values indicating higher likelihood of being distracted. \textit{d} describes how likely a student is to distract other students, with higher values indicating higher likelihood of distracting other students.

\subsection{Fuzzy C-Means Clustering}
With each entry now assigned coefficients, a classification method is needed to interpret the results. FCM clustering is used such that each student entry can be classified according to pre-defined cluster centers whilst retaining membership values. In this way, each student entry can be first classified according to its cluster, then classified again according to its membership to its primary cluster. The number of clusters are determined from the total number of logical coefficient evaluations. A chart of these evaluations is provided in Figure 3. It is highly unlikely to have a student entry that has a low \textit{D} whilst having a high \textit{d}. The results of FCM clustering are shown in Figure 4 where each student entry is assigned a cluster with membership values to all clusters. 

\begin{figure}
\includegraphics[width=\textwidth]{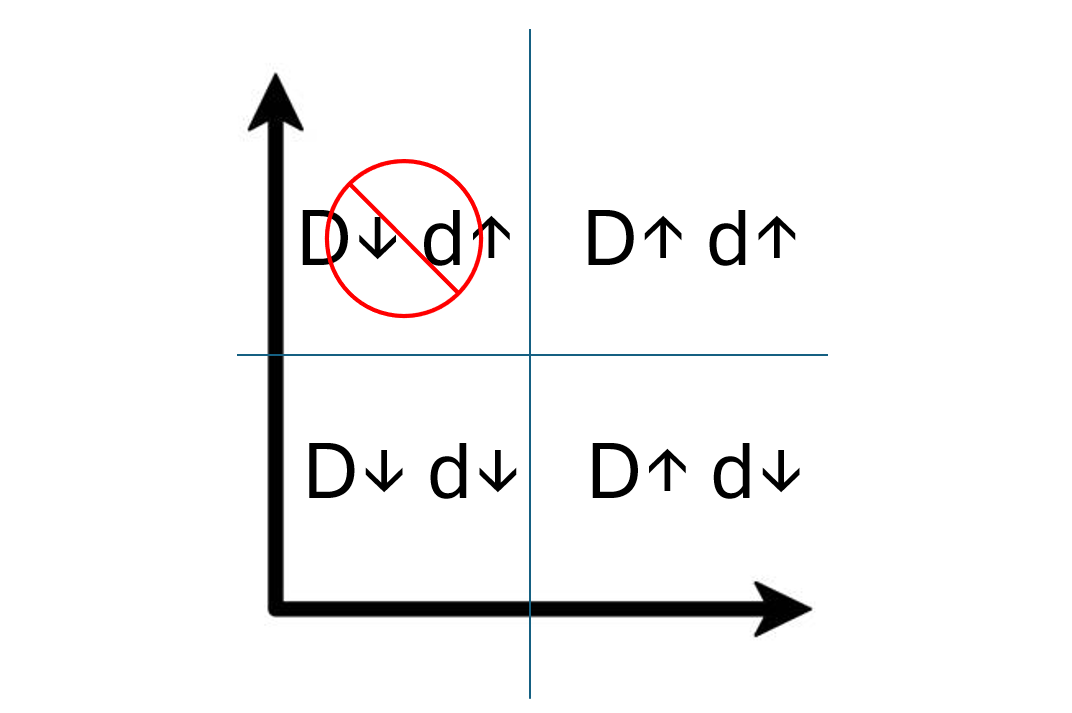}
\caption{Evaluation sectors} \label{fig3}
\end{figure}

\begin{figure}
\includegraphics[width=\textwidth]{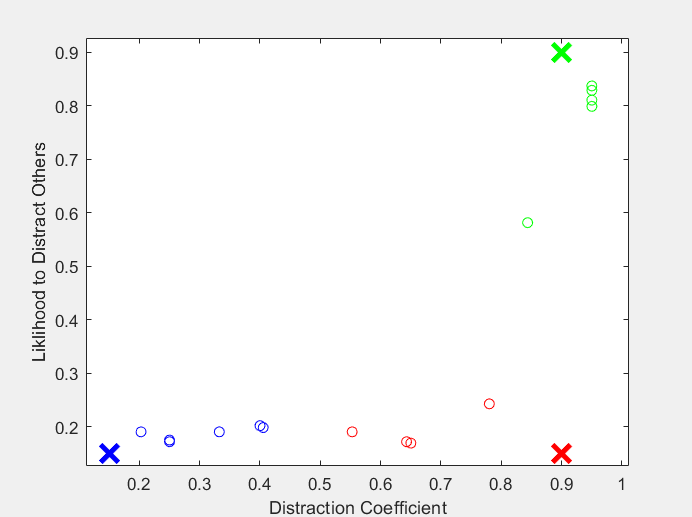}
\caption{FCM clustering results} \label{fig4}
\end{figure}

\subsection{Classification}
At this stage, each student entry has been labeled with a cluster number and cluster membership values. With each student entry, it is plausible to have students who are closely or loosely associated with their cluster centers. Due to the variance in cluster center membership, it is necessary to further classify each student entry. A classification algorithm is used to separate student entries within their respect cluster centers into “High” association and “Low” association. Ultimately, there are six total unique labels possible for each student entry. Figure 5 illustrates the possible labels. For example, “Student 1” may have the classification “Cluster 1, Low Association” which corresponds to a low \textit{D} value and low \textit{d} value that may be greater than the average student entry for “Cluster 1.”

\begin{figure}
\includegraphics[width=\textwidth]{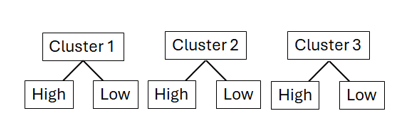}
\caption{The 6 potential labels} \label{fig5}
\end{figure}

\subsection{Seat Assignment}
With each student entry labeled with one of six potential labels, each student entry can be assigned a seat. The algorithm used to determine seating location is a hybrid of sequential and iterative algorithms inspired by the work of \cite{LAI2016780} in solving the “Maximally Diverse Grouping Problem.” Upon completion of the algorithm, each student entry is sorted into groups that best mitigate distraction. The number of groups and the size of each group are defined initially by the teacher. If more seating arrangements are needed, the iterative portion of the algorithm is run again to swap a small number of students creating similar yet new arrangements without losing distraction minimization. It is important to note that distraction minimization is lost after creating many new seating arrangements. In the case of the current study, seating arrangements are cycled monthly, requiring a total of ten unique seating arrangements according to the educational facility’s academic calendar. Minimized distraction would be lost after anywhere between twelve and eighteen subsequent seating arrangements.

\section{Results}
CUB was applied to a modern classroom during the Spring semester. The first ten seating arrangements that CUB produced were evaluated against prior seating arrangements of the same classroom in the same academic year. It was found that the prior seating arrangements labelled as “Good” were nearly identical to the arrangements produced by CUB. One arrangement did not closely match any of the prior arrangements. An example of an evaluation is shown in Figure 6.

\begin{figure}
\includegraphics[width=\textwidth]{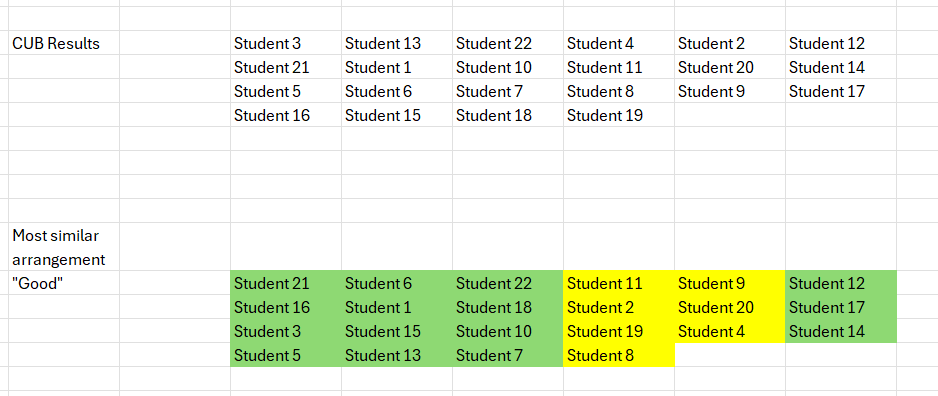}
\caption{A near identical seating arrangement with a "Good" label} \label{fig6}
\end{figure}

\section{Discussion}
CUB produces clustered seating arrangements with high applicability using qualitative individualized information within a small time frame. The produced seating arrangements in the current study are effective enough to be used immediately in their demonstrated classroom. The degree of error is somewhat unknown due to the nature of the problem. With each class having a limited number of seating arrangements based upon the needs of the teacher and the successful evaluation of a seating arrangement based upon the teacher’s qualitative discernment, the loss of minimized distraction is infrequent. As a result, it may be sufficient to establish a threshold for allowed error based upon the goals of the teacher. Another consideration is the opportunity for student behavior to change. As the semester progresses, it may be necessary to adjust survey answers according to changing student behavior to produce applicable seating arrangements. A final consideration would be for the generation of replicated seating arrangements. Establishing a constraint that students in the same group in one arrangement can’t be with each other in the next arrangement would lower the likelihood of duplicate arrangements.
\section{Conclusion}
CUB is a fuzzy logic-based classroom seating arrangement algorithm that produces applicable seating arrangements according to teacher-specified qualitative observations of a student’s tendency to make noise, stay in their seat, and focus on assigned tasks. Although CUB is limited to cluster/group seating arrangements, the produced seating arrangements increase the effectiveness of a teacher’s classroom management. Further investigation can be conducted on non-cluster/group seating arrangements.

\subsubsection{\ackname}
The authors extend their sincere gratitude to the members of the AI Bio Lab at the University of Cincinnati for their insightful conversations, intuitive suggestions, and genuine support that helped realize this work. In particular, the contributions of Wilhelm Louw, Jared Burton, Bharadwaj Dogga, Lohith Pentapalli, Nate Steffen, Magnus Sieverding, Tri Nguyen, Lucia Vilar Nuño, Hugo Henry, and Kaus Shankar are greatly appreciated.

%
%
%
%


\newpage

\printbibliography

@article{HOEKSTRA2023104016,
title = {Teachers’ goals and strategies for classroom seating arrangements: A qualitative study},
journal = {Teaching and Teacher Education},
volume = {124},
pages = {104016},
year = {2023},
issn = {0742-051X},
doi = {https://doi.org/10.1016/j.tate.2023.104016},
url = {https://www.sciencedirect.com/science/article/pii/S0742051X23000045},
author = {Nathalie A.H. Hoekstra and Yvonne H.M. {van den Berg} and Tessa A.M. Lansu and M. Tim Mainhard and Antonius H.N. Cillessen}
}

@inbook{Zadeh2023,
author="Zadeh, Lotfi A.",
editor="Lin, Tsau-Young
and Liau, Churn-Jung
and Kacprzyk, Janusz",
title="Fuzzy Logic",
bookTitle="Granular, Fuzzy, and Soft Computing",
year="2023",
publisher="Springer US",
address="New York, NY",
pages="19--49",
isbn="978-1-0716-2628-3",
doi="10.1007/978-1-0716-2628-3_234",
url="https://doi.org/10.1007/978-1-0716-2628-3_234"
}

@article{LAI2016780,
title = {Iterated maxima search for the maximally diverse grouping problem},
journal = {European Journal of Operational Research},
volume = {254},
number = {3},
pages = {780-800},
year = {2016},
issn = {0377-2217},
doi = {https://doi.org/10.1016/j.ejor.2016.05.018},
url = {https://www.sciencedirect.com/science/article/pii/S0377221716303381},
author = {Xiangjing Lai and Jin-Kao Hao}
}

\end{document}